\documentclass[fleqn,usenatbib]{mnras}

\usepackage[T1]{fontenc}

\DeclareRobustCommand{\VAN}[3]{#2}
\let\VANthebibliography\thebibliography
\def\thebibliography{\DeclareRobustCommand{\VAN}[3]{##3}\VANthebibliography}

\usepackage{graphicx}	
\usepackage{amsmath}	
\usepackage{amssymb}	

\title{Quasi-periodic spicule-like cool jets driven by Alfv\'en pulses}

\author[B. Singh et al.]{
B. Singh,$^{1}$ 
A.K. Srivastava,$^{1}$ \thanks{Corresponding Author: A.K. Srivastava (asrivastava.app@iitbhu.ac.in)}
K. Sharma,$^{1}$
S.K. Mishra,$^{2}$
 and B.N. Dwivedi$^{3}$
\\
$^{1}$ Department of Physics, Indian Institute of Technology (BHU), Varanasi 221005, India \\
$^{2}$Indian Institute of Astrophysics, Kormangala, Bangalore, India\\
$^{3}$RGIPT, Jais Amethi - 229304, India
}

\date{Accepted 2022 January 24. Received 2022 January 23; in original form 2021 November 15}

\pubyear{2021}

\begin{document}
\label{firstpage}
\pagerange{\pageref{firstpage}--\pageref{lastpage}}
\maketitle

\begin{abstract}
We perform a 2.5 dimensional magnetohydrodynamic (MHD) simulation to understand a comprehensive view of the formation of spicule-like cool jets due to initial transverse velocity pulses akin to Alfv\'en pulses in the solar chromosphere. We invoke multiple velocity ($V_{z}$) pulses between 1.5 and 2.0 Mm in the solar atmosphere, which create the initial transverse velocity perturbations. These pulses transfer energy non-linearly to the field aligned perturbations due to the ponderomotive force. This physical process further creates the magnetoacoustic shocks followed by quasi-periodic plasma motions in the solar atmosphere. The field aligned magnetoacoustic shocks move upward which subsequently cause quasi-periodic rise and fall of the chromospheric plasma into the overlying corona as a thin and cool spicule-like jets. The magnitude of the initial applied transverse velocity pulses are taken in the range of 50-90 km $s^{-1}$. These pulses are found to be strong enough to generate the spicule-like jets. We analyze the evolution, kinematics and energetics of these spicule-like jets. We find that the transported mass flux and kinetic energy density are substantial in the localized solar-corona. These mass motions generate $\it in$ 
$situ$ quasi-periodic oscillations on the scale of $\simeq$ 4.0 min above the transition region.
\end{abstract}

\begin{keywords}
(magnetohydrodynamics) MHD; shock waves; waves; Sun: chromosphere, Sun: corona; Sun: oscillations.
\end{keywords}



\section{Introduction}

    Extensive observations and theoretical modeling reveal that the spicule-like jets are frequently and quasi-periodically trigger in the magnetized Sun's atmosphere at disparate spatial and temporal scales \citep[e.g.,][and references cited therein] {2000SoPh..196...79S, 2004Natur.430..536D, 2007Sci...318.1574D, 2010ApJ...722.1644S, 2011Natur.475..477M, 2017NatSR...743147S, 2017SPD....4810403M, 2019NatCo..10.3504L, 2020ApJ...897L...2P, 2021ApJ...913...19M, 2021ApJ...918L..20H}. The solar chromosphere consists of a very complex structuring of the magnetic field, which generates different types of plasma ejecta (e.g., small-scale cool jets, spicules, magnetic swirls, network jets, surges, macrospicules, etc). These features display the existence of various plasma processes within them in the chromosphere of the Sun's atmosphere, e.g., waves, plasma flows, instabilities, etc. These spicule-like jets and other plasma ejecta may also play a significant role in transporting the mass and energy into the overlying region, which results in the localized heating of the quiet-Sun atmosphere \citep[e.g.,][and references cited therein]{2007Sci...318.1574D, 2014Sci...346D.315D, 2011A&A...535A..58M, 2014Sci...346A.315T, 2015A&A...581A.131J, 2018NatAs...2..951S, 2020ApJ...894..155S, 2018A&A...616A..99K, 2019AnGeo..37..891S, 2020ApJ...897L...2P, 2021ApJ...913...19M, 2021ApJ...918L..20H}.\\ 
    
    Recent observations show that various types of spicule-like jets trigger from the quiet chromosphere (e.g., macrospicules, small scale cool jets, network jets, magnetic swirls etc.). These jets are widely detected in the chromospheric Ca II, H$\alpha$, and Na D3 lines at 2000 km above the Sun's surface as a cool, dense and thin plasma structures. These spicules are also rise and fall quasi-periodically above the solar limb \citep[e.g.,][]{1998ESASP.421...19S, 2000SoPh..196...79S,2000A&A...360..351W,2009SSRv..149..355Z,2011Natur.475..477M, 2017ApJ...849...78K,2020ApJ...897L...2P,2021ApJ...913...19M}. Recent high resolution (0.2 arcsec) observations from Swedish Solar Telescope (SST) reveal that most of the spicule-like cool jets appear with their double thread structures and show the bidirectional plasma flows. These structures are identical to the quiet-Sun mottles and active region dynamic fibrils. These spicule-like jets trigger with supersonic speed of $\simeq$ 25 km $s^{-1}$ along the magnetic field lines \citep[e.g.,][]{ 2009SoPh..260...59P, 2010A&A...519A...8M}. They attain the maximum height of 5-7 Mm above the solar surface. Thereafter, they fade away or downfall towards their site of origin. The typical lifetime of these spicule-like jets is about 5-15 min, and width is $\simeq$ 600 km at height between 4 Mm and 10 Mm above the solar photosphere \citep[e.g.,][]{2009SoPh..260...59P}.\\
    
    Recent observations by Hinode/Solar Optical Telescope (SOT), however, show quite different physical properties of the solar spicules \citep[e.g.,][]{2007Sci...318.1574D}. These spicules are classically referred as type II spicules. In the present scenario, the jets which are similar to cool and thin spicules are classically observed on the basis of different morphology referred as type I (i.e. spicules with supersonic velocity of $\simeq$ 25 km $s^{-1}$ observed above the solar limb), and type II spicules \citep[e.g.,][]{2014Sci...346A.315T}. The type II spicules are observed in Ca II H emissions that depict shorter life time of 50-150 s, shorter height of about 4 Mm and smaller diameters of about 200 km. The typical speed of type II spicules is about 50-100 km $s^{-1}$, which is much higher than type I spicules \citep[e.g.,][]{2007Sci...318.1574D, 2009ApJ...705..272R, 2010A&A...519A...8M}. The maximum height of type II spicules observed in the coronal holes is 5 Mm, which appears shorter in the solar active region and quiet-Sun atmosphere. These spicules are not observed over the plages region \citep[e.g.,][]{2007Sci...318.1574D,2012ApJ...746..158J,2017Sci...356.1269M}. The motion and evolution of type II spicules mimic upward flow of chromospheric plasma, which disappears later on. \\
    
    Although there exist various theoretical models to understand the evolution and formation mechanisms of such chromospheric ejecta (e.g., spicule-like cool jets, magnetic swirls, network jets, fibrils, confined surges etc.), yet none of them fully explain their origin \citep[e.g.,][and references cited therein]{2010A&A...519A...8M,2011ApJ...736....9M,2011A&A...535A..58M,2018NatAs...2..951S,2020ApJ...897L...2P,2021ApJ...913...19M}. The non-linear MHD simulation is one of the best possible ways to provide a reasonable understanding of the evolution and dynamics of cool jets. These models may be useful to explain the spicule-like jets triggered by the action of pulses, which are based on the energy deposition locally in the solar chromosphere \citep[e.g.,][and references cited therein]{2000SoPh..196...79S, 2010A&A...519A...8M, 2021ApJ...913...19M}. The deposition of energy and formation of spicule-like jets have been studied invoking the launch of impulsive velocity perturbation or gas pressure perturbation in the lower solar chromosphere \citep[e.g.,][and references cited therein]{1982SoPh...75...99S,1990ApJ...349..647S,1993ApJ...407..778S,  2010A&A...519A...8M, 2019AnGeo..37..891S}. These perturbations result into the magnetoacoustic shocks, which eventually eject the chromospheric plasma through transition region into the upper solar atmosphere. These shocks may exhibit after effects of the heating scenario in the chromosphere. The shock-generated model provides the main idea about an evolution of spicule-like jets and other chromospheric ejecta such as network jets, macrospicules, cool jets, swirls etc \citep[e.g.,][and references cited therein]{1982SoPh...75...35H, 2007ApJ...666.1277H, 2010A&A...519A...8M, 2011A&A...535A..58M, 2011ApJ...736....9M, 2019AnGeo..37..891S, 2021ApJ...913...19M}. \\
    
    \citet[][] {1982SoPh...75...99S} have reported a model based on slow-mode waves similar to the shock model describing the generation of spicules. These plasma ejecta show a similar behaviour as of thin, cool and small-scale jets in the lower solar chromosphere. However, the velocity perturbation and gas pressure perturbation show one of the possible origin of slow magnetoacoustic waves and associated shocks to trigger such cool jets. These perturbations interact with granular motions, which enable the small-scale magnetic reconnection at different heights in the chromosphere. As a result, the energy is released with the localized brightening that triggers the spicule-like jets. The magnitude of reconnection and its height describe the evolution of various physical processes. For instance, the reconnection in the chromosphere describes the evolution of magnetoacoustic shock. If the reconnection occurs in the inner corona, the Alfv\'en wave may play a significant role in formation of jets \citep[e.g.,][]{2008ApJ...683L..83N, 2013ApJ...770L...3K}. The generation of spicules by Alfv\'en waves has also been discussed by \citet[][]{1982SoPh...75...35H} in the lower solar atmosphere. The Alfv\'en waves excited by the random perturbations of $\simeq$ 1 km $s^{-1}$ launched at the solar photosphere may be responsible for the generation of spicule-like jets \citep[e.g.,][]{1982ApJ...257..345H, 1999ApJ...514..493K}. In this model, Alfv\'en waves exhibit coupling between the generated fast magnetoacoustic shocks/waves and the formation of spicules. This non-linear coupling may lift up the plasma of transition region, and generate the spicule-like small-scale jets. An another mechanism for the production of spicules is observed when ion-neutral collisions damp the high-frequency Alfv\'en waves in the solar atmosphere \citep[e.g.,][]{2003A&A...406..715J}. These models may describe several observational characteristics of spicule-like cool jets. Recently, the numerical models and observations have shown that macrospicules are also triggered impulsively in the solar atmosphere \citep[e.g.,][] {2015ApJ...808..135B, 2021ApJ...913...19M}. The two-fluid spicule model proposed by \citet[][]{2017ApJ...849...78K} provides the triggering mechanism by the action of velocity pulse in the vertical direction of the solar atmosphere. There are also several two-fluid simulations available describing the formation of the waves and transients in the solar chromosphere \citep[e.g.,][]{2013ApJ...767..171S, 2019A&A...630A..79P}. These models consist of non-ideal terms such as heat conduction, cooling and radiation, which may play a significant role in the evolution of the spicules \citep[e.g.,][]{1990ApJ...349..647S, 1993ApJ...407..778S, 2007ApJ...666.1277H}. \\ 
   
   In the present paper, we perform a 2.5D numerical simulation  of ideal, non-linear MHD equations to trigger the spicule-like cool and small-scale jets due to evolution of randomly generated Alfv\'en pulses in the solar atmosphere. These pulses are implemented between 1.5 and 2.0 Mm height of the model chromosphere with realistic temperature, pressure, and stratification. These random Alfv\'en pulses are strong enough in the lower-chromosphere (50-90 km $s^{-1}$). The strong amplitude Alfv\'enic perturbations are observed in the solar chromosphere even before and provide a compelling ground to believe that they can be in action to produce the chromospheric dynamical processes \citep[e.g.,][]{2007Sci...318.1574D, 2011Natur.475..477M}. These Alfv\'en pulses while acting as a driver, eventually release the energy in the localized region. This further results into the shock waves and mimic a variety of spicule-like cool jets in the gravitationally stratified solar atmosphere. Although these spicule-like jets trigger from the solar chromosphere, yet their feature can also be observed in the overlying region in the solar atmosphere. Here, we investigate the kinematical properties, energetics and triggering mechanism of such spicule-like jets and their connection to the Alfv\'en pulses and the field aligned magnetoacoustic shock waves. The present paper is organized as follows. The basic MHD model of the solar atmosphere and the equilibrium configuration are given in Sect. 2. The simulation results are presented in Sect. 3. This work is concluded with a discussion and summary in Sect. 4.\\

\begin{figure*}  
\centering
\centerline{\includegraphics[width=0.5\linewidth]{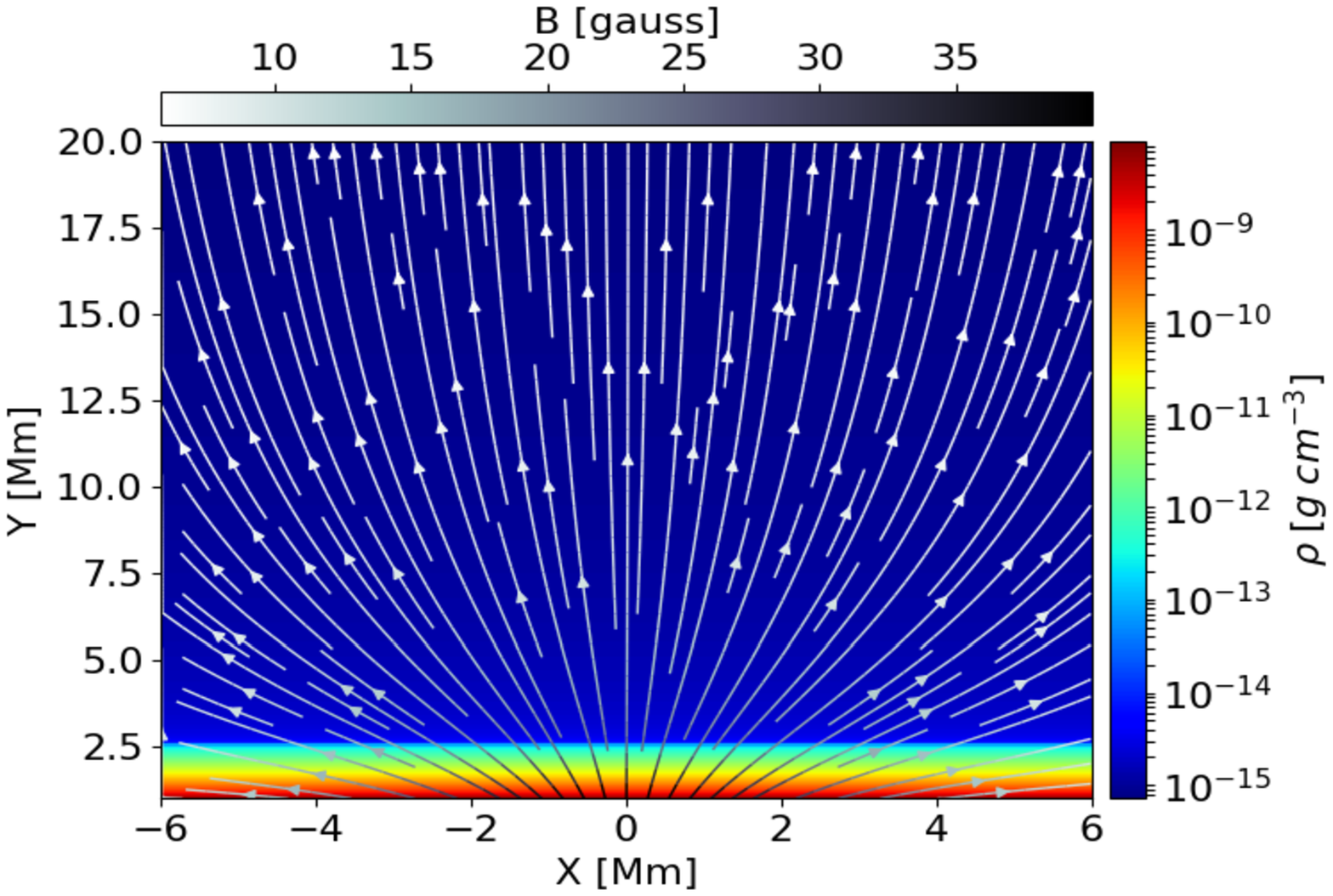}
  \includegraphics[width=0.45\linewidth]{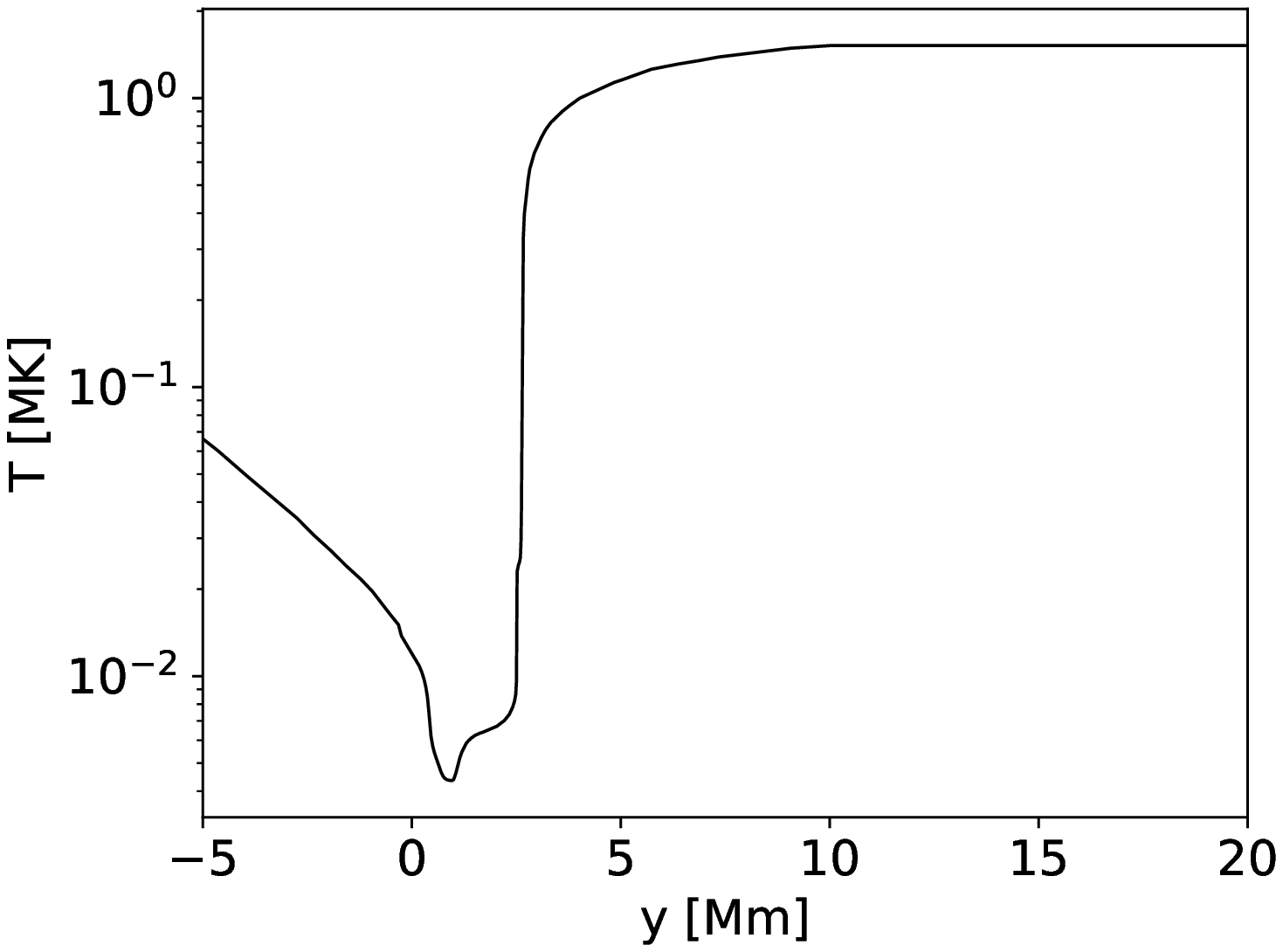}}
\caption{Left: The open expanding magnetic field lines are shown in the stratified solar atmosphere at equilibrium. The source magnetic pole is kept at -3.0 Mm below the boundary of the solar photosphere at x=0 Mm. The magnetic pole strength is taken as $\simeq$ 560 G. Right: The temperature profile w.r.t. the vertical height of the solar atmosphere is displayed as per the model given by \citet{2008ApJS..175..229A}.}
\end{figure*}

\section{Basic model of the solar atmosphere and methods}
\subsection{The ideal MHD system}

We consider the conservative form of the basic ideal MHD equations to model the generation of spicule-like jets due to the initial evolution of impulsive non-linear Alfv\'en pulses in the magnetized and gravitationally stratified solar atmosphere \citep[e.g.,][]{2007jena.confR..96M,2019AnGeo..37..891S,2020ApJ...894..155S}.
\begin{equation}
    \frac{\partial\varrho}{\partial t} + \nabla\cdot(\varrho\textbf{V}) = 0  
\end{equation}

\begin{equation}
     \varrho\frac{\partial\textbf{V}}{\partial t} +\varrho(\textbf{V}\cdot\nabla)\textbf{V} = -\nabla p + \frac{1}{\mu}(\nabla\times\textbf{B})\times\textbf{B} +\varrho\textbf{g}
\end{equation}

\begin{equation}
  \frac{\partial\textbf{B}}{\partial t} = \nabla \times (\textbf{V} \times\textbf{B}),\hspace{1cm}    \nabla\cdot\textbf{B} = 0  
\end{equation}

\begin{equation}
  \frac{\partial p}{\partial t} + \nabla \cdot (p\textbf{V}) = (1- \gamma) p \nabla\cdot\textbf{V}  
\end{equation}
\begin{equation}
    p= \kappa_{B} \frac{\varrho T}{m}
\end{equation}

The symbols $\textbf{V}$, and $\textbf{B}$ of the above set of ideal MHD equations are expressed as plasma velocity and magnetic field strength respectively. The symbol '$\textbf{g}$' represents the gravitational acceleration taken as 27400.0 cm $s^{-2}$ which is average value of the solar photosphere. The constants '$\mu$' and '$k_{B}$' depict the magnetic permeability and Boltzmann's constant respectively. The symbol '$\varrho$' denotes the mass density and 'p' depicts thermal pressure. The symbol '$\gamma$' represents specific heats ratio taken for the three degree of freedom i.e., equal to 5/3. The symbol '$ m $' denotes the mean mass of the particle present in the numerical domain of the present model that is equal to 1.24. The 'T' denotes plasma temperature in the model solar atmosphere.  In this paper, we do not consider the velocity of the background plasma flows. We also do not implement non-ideal effects such as magnetic diffusivity, dissipative effect, resistivity, viscosity, plasma cooling and /or heating. We aim to model, triggering, kinematics, and energetics of the spicule-like jets. The embedded magnetic field lines ('$\textbf{B}$') hold the divergence free condition and its divergence is nearly equal to zero. \\

\subsection{Equilibrium model configuration of the solar atmosphere} 
 \label{S-Equli}

Initially, we consider a realistic magnetized solar atmosphere supporting the region above the convection zone to the inner solar corona (i.e., 20 Mm). The considered solar atmosphere is in static equilibrium (i.e., $V_{e}$ = 0). This model holds good for the current-free magnetic field and also approximated with the force-free medium initially. Therefore, the initial condition of magnetic field is expressed as\\
\begin{equation}\\ \\
\nabla \times \textbf{B} = 0 , \hspace{2cm} (\nabla \times \textbf{B}) \times \textbf{B} = 0
\end{equation}

 The equilibrium magnetic field configuration \textbf{B} is  estimated as follows with its components \citep[e.g.,][]{1985ApJ...293...31L, 2019AnGeo..37..891S}:\\

\begin{equation}\\ \\
   B_{x} = \frac{-2S(x-a)(y-b)}{((x- a)^2 + (y - b)^2)^2}
\end{equation}
\begin{equation}\\ \\
    B_{y} = \frac{2S(x-a)^2 - S((x-a)^2 + (y-b)^2)}{((x- a)^2 + (y - b)^2)^2}
\end{equation}

Here 'S' represents the magnetic pole strength while constants 'a' and 'b' collectively describe its position in the embedded solar atmosphere. The eqs (7)-(8) are specified for the magnetic field in the embedded solar atmosphere. The '$B_{x}$' indicates the component of the magnetic field in the horizontal direction while '$B_{y}$' indicates the magnetic field component in y-direction. The other horizontal direction is the z-direction which is out of the plane to x-y. The 'z' is an ignorable horizontal direction. In the present model, we fix the vertical position of the magnetic pole below the photospheric boundary (i.e., = -3.0 Mm) at the center of the simulation domain (i.e. 0 Mm on the horizontal direction). The spicule-like jets trigger from the chromosphere (between 1.5 and 2.0 Mm) along the open magnetic field lines in the model solar atmosphere. The magnetic field lines are originated from the convection zone (i.e., -3.0 Mm). There is no significant dynamics taking place below the photospheric boundary. Therefore, we fix the lower boundary at the chromosphere (i.e., at 1.0 Mm) to optimize the computation time. In this way, the open magnetic field lines are visualized as expanding lines from the solar chromosphere (i.e., 1.0 Mm) to the inner corona up to 20 Mm (see Fig. 1, left panel). The magnetic field decreases with the height in the solar atmosphere (Eq. 8). It is not exactly the exponential decay, but it is essentially the decay of magnetic field algebraically following equation 8 (Fig. 1, left panel).\\  

\begin{figure*} 
\centering
\centerline{\includegraphics[width=0.8\textwidth, clip=]{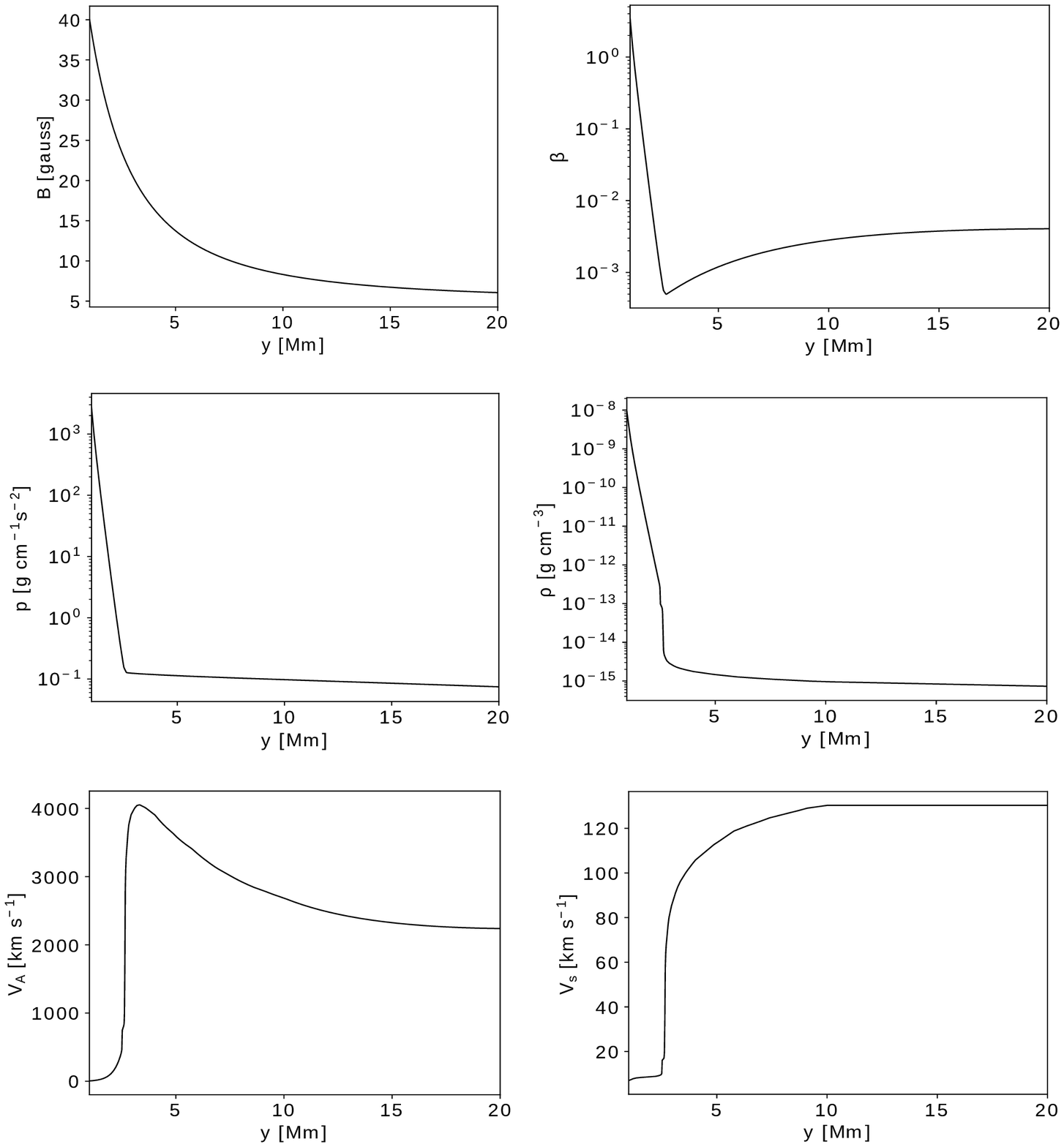}}        
\caption{The magnetic field (top-left), plasma beta (top-right), pressure (middle-left), mass density (middle-right), Alfv\'en  speed (bottom-left), and sound speed (bottom-right) w.r.t. height (y) are displayed in the model solar atmosphere.}
\label{fig}
\end{figure*}

\begin{figure*}   
\centering
\includegraphics[width=1.0\textwidth,clip=]{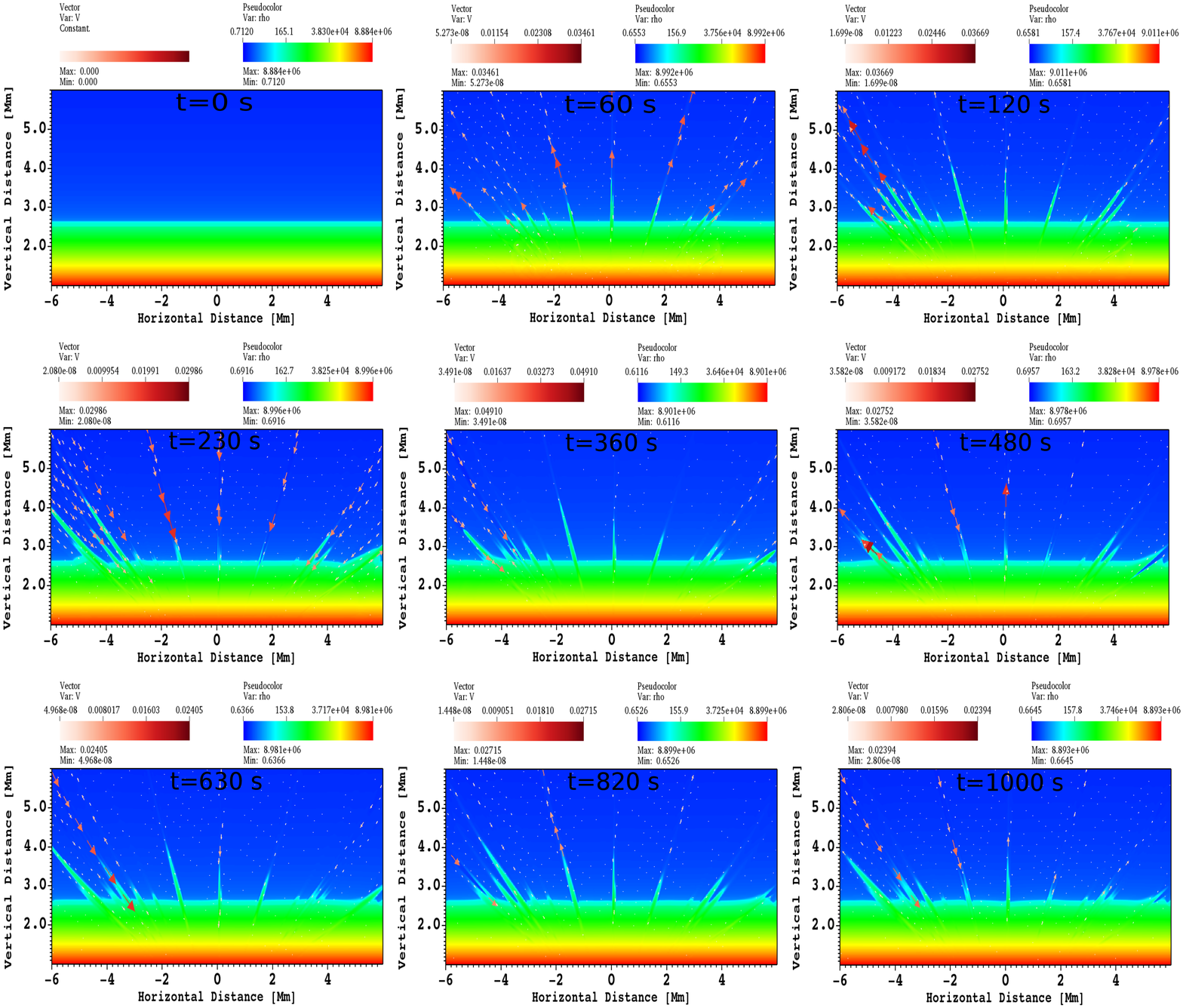}
\caption{The normalized density map of the evolution of different spicule-like jets are shown in this figure. These jets are triggered due to 15 different amplitudes of the $V_{z}$ pulses. The transverse pulses mimic the Alfv\'en pulses which are applied in the chromosphere between 1.5 and 2.0 Mm. The amplitude of these pulses lie between 50 to 90 km $s^{-1}$. They generate the field aligned magnetoacoustic shocks followed by the mass motions exhibiting the features of the spicule-like jets. The normalized velocity vectors are over-plotted on each density maps between t=0 s and t=1000 s.}
\label{}
\end{figure*}

The realistic temperature variation $T_{e}($y$)$ is given in Fig.~1 (right-panel), which is derived by \citet{2008ApJS..175..229A} using observed line profiles and also described in various recent literatures, e.g., \citet[][]{2012A&A...537A..96K, 2019AnGeo..37..891S, 2020ApJ...894..155S}. Fig. 1 (right panel) shows the typical value of the temperature (T) with height (y). Its value at y = 0.5 Mm is 5700 K, which is the photosphere. On moving towards the higher altitudes from the solar photosphere at 0.5 Mm, it gradually decreases and attains its minimum value 4350 K at y = 0.95 Mm. It increases sharply, and attains 432912 K temperature at the height 2.7 Mm, which is the transition region. Thereafter, T(y) sharply increases up to the inner corona and finally attains a mega-Kelvin temperature. \\

The background of the stratified solar atmosphere subsequently holds the hydrostatic equilibrium condition. Therefore, the pressure gradient force is balanced by the gravity force which makes the medium approximately force-free. This can be written as follows: 
 
\begin{equation}\\ \\
- \bigtriangledown p + \rho \textbf{g} = 0
\end{equation}\\
 We determine the equilibrium thermal pressure and mass density w.r.t. vertical direction (y) in the hydrostatic equilibrium of the initial solar atmosphere which is as follows \citep[e.g.,][] {2019AnGeo..37..891S, 2020ApJ...894..155S}:\\
 
\begin{equation}\\ \\
p_{e}(y) = p_{ref} exp\left(- \int_{yref}^{y}\frac{dy'}{\Lambda(y')}\right), \hspace{0.5cm}        \rho_{e}(y) = \frac{p_{e}(y)}{g\Lambda(y)}
\end{equation}
where
\begin{equation}\\ \\
\Lambda(y) =\frac{k_{B} T_{e}(y)}{\hat{m}g}
\end{equation} 
Here, $P_{ref}$ is the thermal pressure at reference level $y_{ref}$.\\

The thermal pressure and equilibrium mass density profiles w.r.t. height (y), determined by the above equations (Eqs. 10-11), are shown in the solar atmosphere in Fig. 2, middle panels. We notice that plasma pressure (p) and equilibrium mass density $(\rho)$ at 1.5 Mm (i.e., chromosphere) are 64.0 g $cm^{-1}s^{-2}$  and 1.537 $\times$ $10^{-10}$ g $cm^{-3}$ respectively. They fall at the higher heights and attain 0.126 g $cm^{-1}s^{-2}$  and 4.6 $\times$ $10^{-15}$ g $cm^{-3}$ values respectively at the transition region (i.e., at y = 2.7 Mm).\\

We set an appropriate value of source magnetic pole strength (S) as $\simeq$ 560 G at (0, -3) Mm in the convection zone below the photospheric boundary. The vertical background magnetic field of 5 Gauss is added to create a strongly magnetized quiet-Sun atmosphere. This field strength mimics the quiet-Sun magnetic network field. The magnetic fields smoothly extend from the convection zone to the inner corona. The magnetic field strength at y = 1.5 Mm (i.e., chromospheric region) is 32.6 Gauss. It continues to fall with the height (y) and attains 22.23 Gauss at y = 2.7 Mm, which is the solar transition region (Fig. 2, top-left panel; Eq. 8). These magnetic field values represent strongly magnetized quiet chromosphere, TR, and corona \citep[e.g.,][]{2006ApJ...646.1421D}. Fig. 2 (top-right) panel shows the variation of plasma-beta with height (y) in the solar atmosphere. Its value varies from 0.12 at y = 1.5 Mm i.e., solar chromosphere  to 0.0005 at y = 2.7 Mm (i.e., transition region).\\

The variation of Alfv\'en speed and sound speed w.r.t. height in the solar atmosphere are shown in Fig. 2 (bottom panels). The  Alfv\'en speed is determined by $V_{A} = \sqrt{\frac{B^2}{\mu\rho}}$, which is 3278.7 km $s^{-1}$ at the solar transition region (y = 2.7 Mm). It decreases in the lower part of the solar atmosphere and attains the value 26.3 km $s^{-1}$ at the chromosphere (i.e., at y = 1.5 Mm). The value of sound speed estimated by $V_{s} = \sqrt{\frac{\gamma p}{\rho}} $ at the transition region (2.7 Mm) is 67.6 km $s^{-1}$. At the solar chromosphere (1.5 Mm), It attains the value 8.3 km $s^{-1}$. The smooth variation of these background physical parameters across the various layers of the magnetized and gravitationally stratified solar atmosphere shows the longitudinal structuring, and exhibits a realistic model atmosphere. This model atmosphere is further utilized to trigger the spicule-like cool jets to understand their physical properties.  \\

\begin{figure*}     
\centering
\includegraphics[width=0.6\textwidth,clip=]{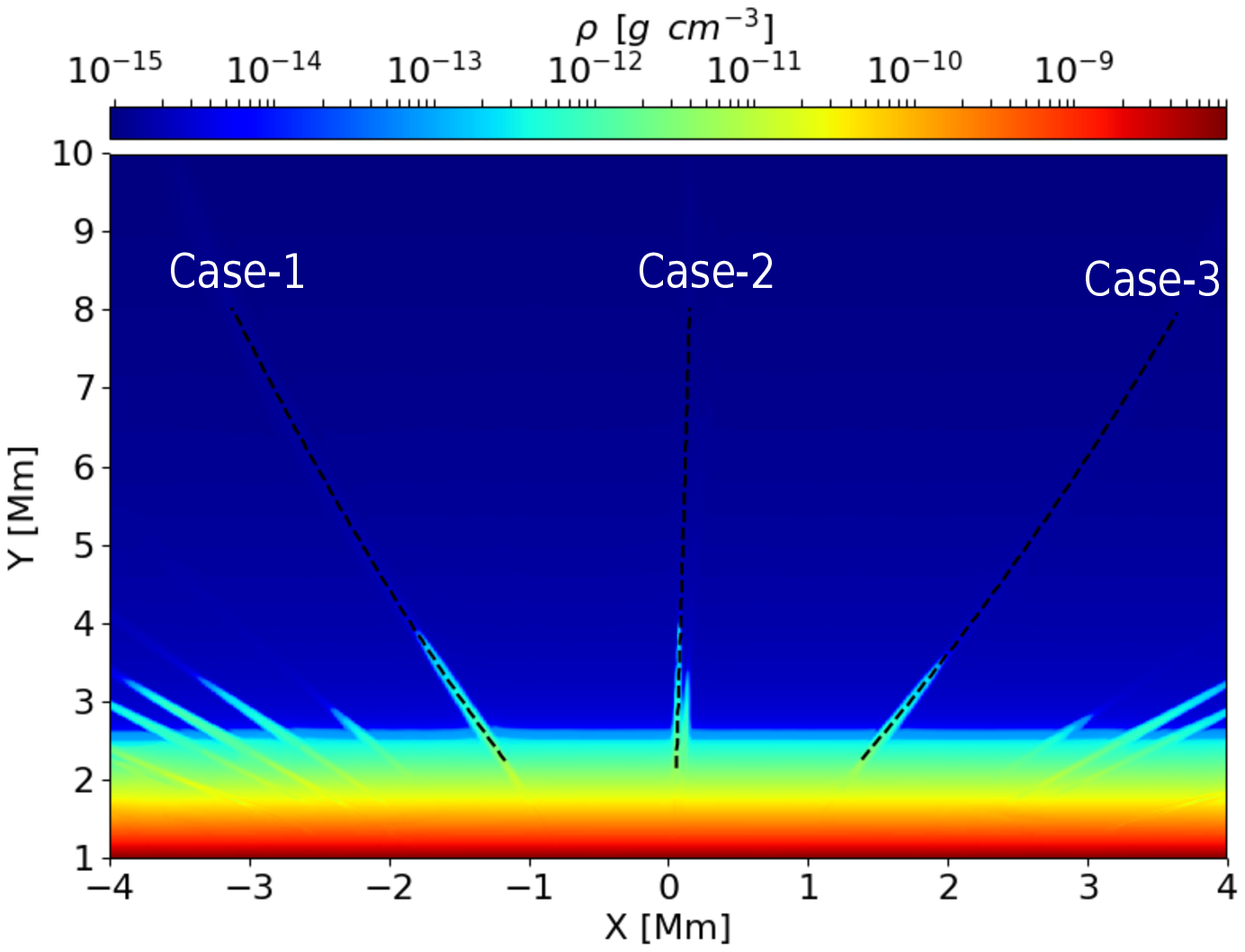}
\includegraphics[width=1.0\textwidth, clip=]{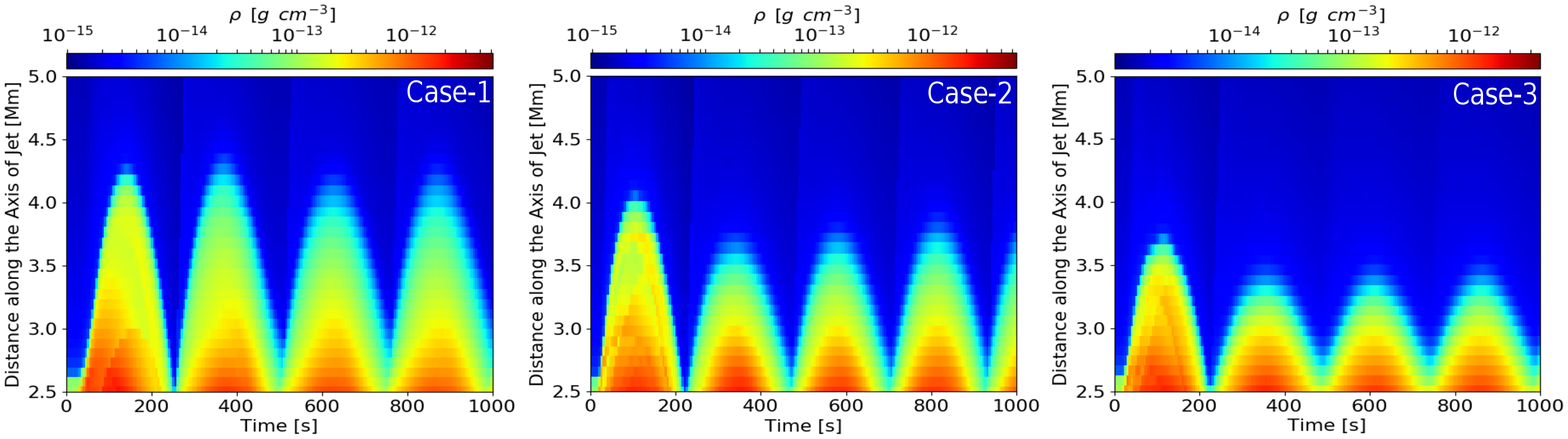}
\caption{Top Panel: The logarithm of mass density ($\rho$) in g cm$^{-3}$ with three dotted slits (case-1, case-2, case-3) is displayed. We estimate the distance-time diagrams over the chosen paths. Bottom Panel: The distance time diagrams corresponding to the chosen paths exhibit quasi-periodic rise and fall of the spicule-like cool jets.}
\end{figure*}

\subsection{Numerical methods}
We solve the set of ideal MHD equations numerically using Godunov-type PLUTO code and implement the initial conditions for the generation of the spicule-like jets. We fix the localized solar atmosphere of 12 Mm (-6, 6) span in the horizontal direction (x), and 20 Mm (1, 21) span in the vertical direction (y). We have divided the horizontal span into 1536 equal cells as static uniform grid. The vertical span is divided into 512 equal cells from 1 to 5 Mm as uniform static grid, and 1024 equal cells from 5 to 21 Mm as the stretched grid. Therefore, the spatial resolution of simulation domain with static uniform grid is 7.812 km per numerical cell in the horizontal as well as vertical direction, while its spatial resolution with static stretched grid is 15.625 km per numerical cells in the vertical direction.\\

We impose all the four boundary conditions of simulation domain so that the boundary keeps maintaining the equilibrium values of all the plasma quantities. We implement such user-defined boundaries as it is the simplest procedure of having reasonably well working boundaries in the presence of gravity in the system. We do not set the reflecting boundaries for velocity because it does not work well with gravity. As having boundary conditions working well with gravity is a formidable task, we implemented the above mentioned boundary condition at the first instance that is working appropriate in the present scientific context and in setting the gravitationally stratified atmosphere to understand its dynamical plasma processes. \\

We have used the appropriate stretched grids in the top-zone of the numerical box. Most of the signals related to the Alfv\'en perturbations lie in the chromosphere itself, and very weak velocity perturbations reached up to the overlying part. It should be noted that the region above 5 Mm are filled with stretched grids. Since, the grids are stretched in an appropriate manner and number in the model corona (top-zone of the numerical domain), the incoming signals are diffused near the top boundary. The use of a min-mod limiter also keeps this aspect working in place. This restricts the potential reflection of the dynamics from the top-zone of the simulation box. Therefore, we do not have any significant dynamics in the system specially in the top part of the numerical simulation box. The region-of-interest is around TR, where most of the energy exchange takes place and field-aligned magnetoacoustic shocks are generated and propagated up. This further causes the quasi-periodic rise and fall of the spicule-like jets up to 4 Mm height.\\

This simulation box is constructed/embedded as the realistic solar atmosphere. The PLUTO code solves the non-linear MHD equations with double precision arithmetic of the conservative system. In the present modeling, the flux computation is solved by Roe solver which is linearized Riemann solver. It solves using the characteristic decomposition of the Roe matrix \citep[cf.,][]{2007jena.confR..96M}. We set the Courant-Friedrichs-Lewy (CFL) number equal to 0.25 and select the third order Runge-Kutta time integration. We fix the temporal resolution equal to 5 sec. We run the simulation up to 1000 s using multiple passage interface (MPI) for the parallel calculations. \\ 

\subsection{Perturbation} 
 
Initially at t= 0 s, we launch the $V_{z}$ perturbations mimicking the Alfv\'en pulses in the embedded realistic solar atmosphere which are responsible for the evolution of a variety of spicule-like cool jets. The Gaussian form of the velocity pulse is expressed as follows:

\begin{equation}\\ \\
V_{z} = A_{v} \times exp \left(-\frac{(x - x_{0})^2 + (y - y_{0})^2}{w^2}\right) 
\end{equation} \\
Here, $A_{v}$ represents the amplitude of the pulse, $ x_{0}$ and  $y_{0} $ collectively depict its position, $w$ represents the width of the Gaussian pulse which is equal to 20 km. In this simulation, we generate 15 random set of magnitude of transverse velocity pulse ($V_{z}$) between 50 and 90 km $s^{-1}$, applied vertically ($ y_{0}$) between 1.5 and 2.0 Mm, and horizontally ($x_{0}$) between -5.0 and 5.0 Mm. These amplitude of pulses (50-90 km $s^{-1}$) are strong enough to trigger the spicular-like jets from the chromosphere along the magnetic field lines. The impulsive, non-linear $V_{z}$ pulses are chosen keeping the view that strong amplitude transverse perturbations and associated Alfv\'en wave motions have already been observed in the solar atmosphere \citep[e.g.,][]{1999A&A...349..636T, 2007Sci...318.1574D, 2011Natur.475..477M}. Such transverse pulses may evolve in highly localized regions (i.e., even over the un-resolved spatial scales) in the solar chromosphere which can further perturb the magnetic fields and can launch Alfv\'en waves/pulses non-linearly \citep[e.g.,][]{2014C&C...43..42W}. In the presence of curved magnetic field lines in pure 2-D simulations, the horizontal velocity perturbations produce essentially the magnetoacoustic-gravity waves propagating along the magnetic field lines in the vertical direction \citep[e.g.,][]{2013MNRAS.434.2347J}. However in the present 2.5 D simulation, the pure Alfv\'en perturbations are generated by the implementation of velocity perturbations in the z-direction. The transverse waves propagate in the plane of the magnetic field (x-y plane), however it's polarization (velocity perturbations) lies normal to this plane in the ignorable horizontal direction (i.e., z-direction). \\

\begin{figure*}      
\centering
\includegraphics[width=1.0\textwidth]{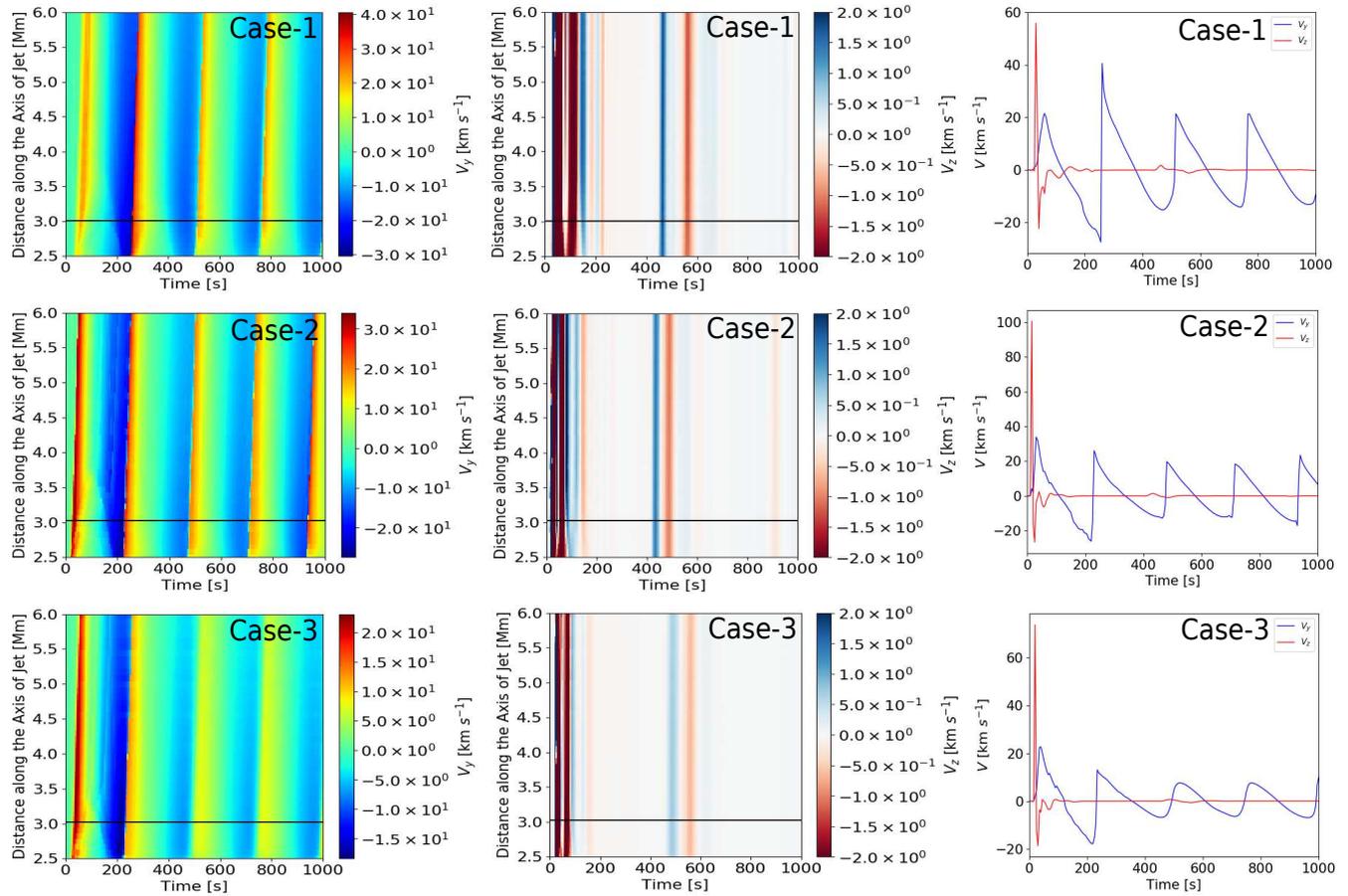}
\caption{Left column: The distance time diagram of the upward velocity ($V_{y}$) is shown along the chosen paths where jets are formed. Middle Column: The distance time diagram of $V_{z}$ is displayed, which is scaled from -2.0 km $s^{-1}$ to 2.0 km $s^{-1}$. These $V_{y}$  and $V_{z}$ are taken corresponding to the slits drawn on Fig. 4 (black-dotted lines) for three cases, i.e., case-1, case-2 and case-3. Right column: The temporal evolution of vertical velocity $V_{y}$ (blue curve) and transverse velocity $V_{z}$ (red curve) corresponding to the horizontal black slits on the left and middle columns respectively taken at y=3.0 Mm.}
\end{figure*}

\begin{figure*}     
\centerline{\includegraphics[width=0.95\textwidth,clip=]{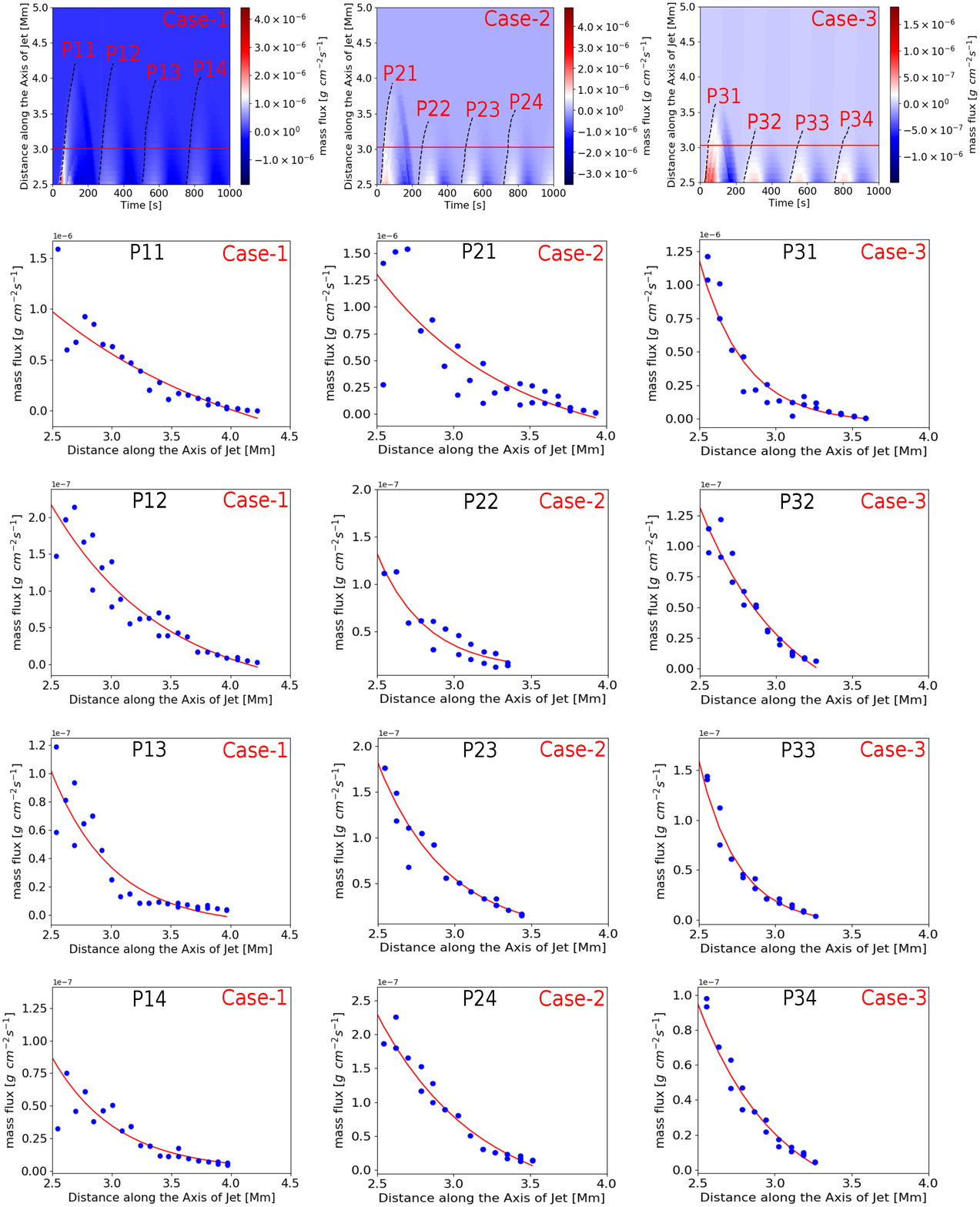}}
\caption{The distance time diagram of the mass flux during the evolution of spicule-like jets is shown in the solar atmosphere (top-row). These diagrams correspond to the slits on Fig. 4, upper panel (black-dotted lines) for the three cases (i.e., case-1, case-2 and case-3). The over-plotted black-dashed lines on each map (top-row) show the paths where we estimate the mass flux w.r.t. the length of the jet. The spatial distributions of the mass flux of various jets are shown in rows 2 to 5 during their maximum elongation in the upward direction.}
\end{figure*}

\begin{figure*}    
\centerline{\includegraphics[width=0.95\textwidth,clip=]{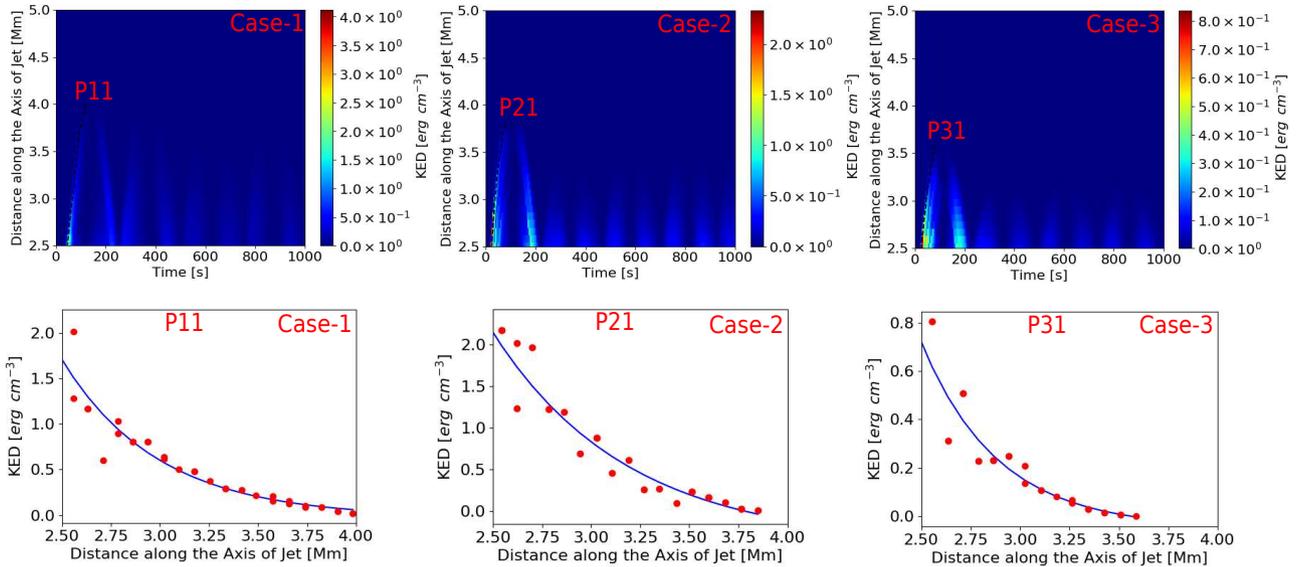}}
\caption{The distance time diagrams of the kinetic energy density (KED) are shown in the top panels during the evolution of spicule-like jets in the solar atmosphere. These diagrams correspond to the slits as shown on Fig. 4 upper-panel (black-dotted lines) for three cases (i.e., case-1, case-2 and case-3). The over-plotted black-dashed line on each maps (top row) show the path where we estimate the KED  w.r.t. the length of the first jet. The bottom row displays the spatial distributions of KED associated with the first spicule-like jets (P11, P21, P31) during their maximum elongation in the upward direction.} 
\end{figure*}
  
\begin{figure}  
\centerline{\includegraphics[width=0.4\textwidth]{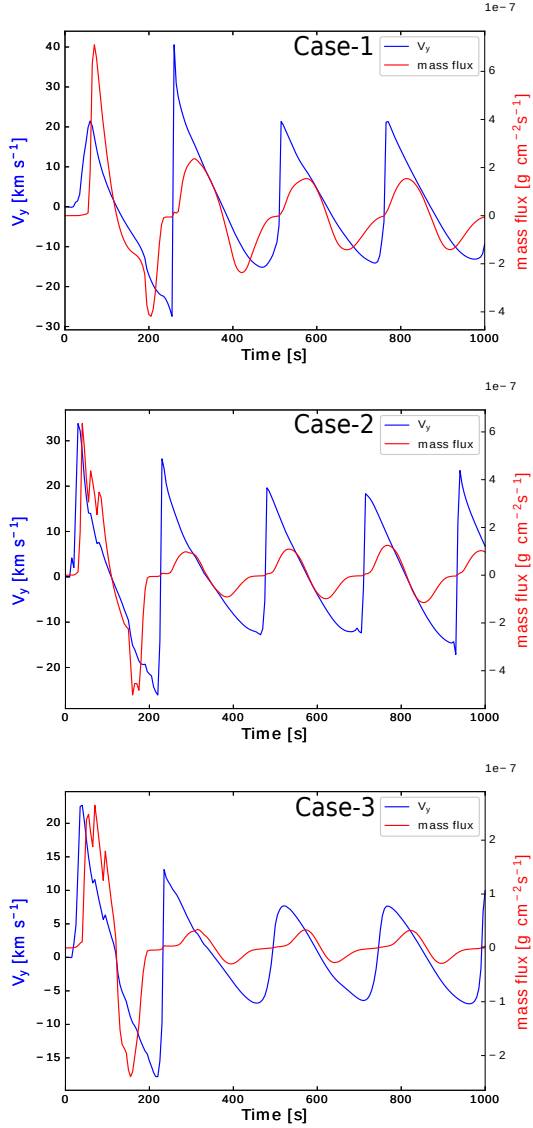}}
\caption{The temporal evolution of the upward velocity $(V_{y})$ in the solar atmosphere (blue curve) is plotted corresponding to the horizontal black slits as shown on the left-most panels of Fig. 5. These slits are drawn there at y = 3.0 Mm. The temporal evolution of the mass flux (red curve) corresponding to the black dotted horizontal line as chosen on Fig. 6 (top row) at y = 3.0 Mm is also compared for the three cases (i.e., case 1, 2 and 3).}
\end{figure}

\begin{figure*}    
\centerline{\includegraphics[width=0.8\textwidth,clip=]{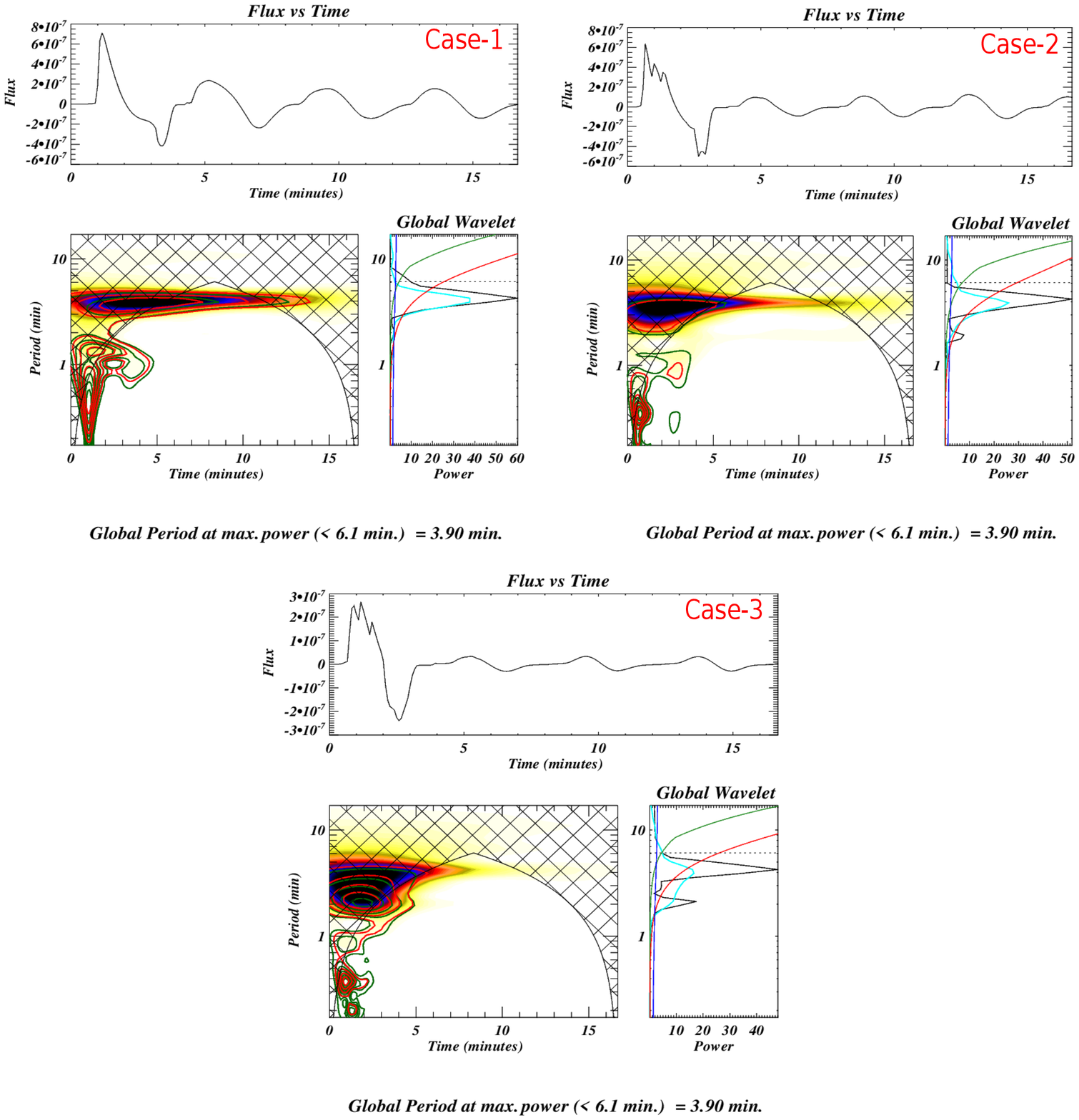}}
\caption{The wavelet analysis of the temporal variation of the mass flux launched by various spicule-like jets along each chosen paths (i.e., case 1-3). The mass flux variation is estimated at the coronal height of 3.0 Mm.}
\end{figure*}

\section{Results} 
In 2.5D regime, the $V_{z}$ perturbations are confined in a very thin spatial region generating large amplitude (non-linear) Alfv\'en waves/pulses. The applied random $V_{z}$ pulses akin to the non-linear Alfv\'en pulses launch the spicule-like thin and cool jets from the chromosphere to the inner corona (Figs. 3-4). The stacked mosaic diagram (Fig. 3) shows the variation of log $\rho$ during the evolution of spicule-like jets at various spatio-temporal scales. The initial evolution of such transverse Alfv\'en pulse later creates shocks and mass motions (Fig. 5). These spicule-like jets carry chromoshpheric energy and mass into the inner corona along with the open magnetic field lines (Figs. 6-7). They are evolved quasi-periodically in the solar atmosphere (Figs. 4, 8-9), as typically observed \citep[e.g.,][]{2004Natur.430..536D, 2007Sci...318.1574D, 2010A&A...519A...8M}. As the transverse $V_{z}$ pulses are launched below the transition region, the chromospheric plasma gets essential velocity disturbance at their base which later exhibits the signature of magnetoacoustic shock waves (cf., right column of Fig. 5). These magnetoacoustic shocks grow into the upward direction along with the expanding field lines. The shocks create the under pressure regions below and cool jets are driven upward in the solar atmosphere. In the non-linear regime, these Alfv\'en pulses transfer energy to the field aligned magnetoacoustic perturbations and plasma flows due to ponderomotive forces \citep[e.g.,][]{2015A&A...577A.126M}. Therefore, the Alfv\'en pulses play a role in the impulsive origin of spicule-like cool plasma ejecta. \\

Previously, the implementation of the Alfv\'en pulses of the order of 1 km $s^{-1}$, once applied on the photosphere, trigger the spicules as reported by \citet[][]{1982ApJ...257..345H}. In our present model, we have applied the random non-linear Alfv\'en pulses in the chromosphere between 1.5 and 2.0 Mm with the amplitude of 50 to 90 km $s^{-1}$. In the initial work by \citet[][]{1982ApJ...257..345H}, the coupling of applied Alfv\'en perturbations to the magnetoacoustic shocks was responsible for the generation of the spicules in the lower solar atmosphere. In the present model, the implemented transverse perturbations generate the pure Alfv\'en pulses that do not couple with the magnetoacoustic waves. However, they transfer the energy to the field-aligned magnetoacoustic perturbations that evolve into the shocks. This further generates the field aligned mass motions in the form of spicule-like cool jets. The Alfv\'en waves consist of kinetic energy and magnetic energy, which is transferred to the plasma in the non-linear regime. In the frame-work of single-fluid magnetohydrodynamics (MHD), as explained above, the ponderomotive force creates field-aligned mass transport following the magnetoacoustic shocks which are eventually moving ahead \citep[e.g.,][and references cited therein]{1982ApJ...257..345H, 1997SoPh..175...93N, 1999JPlPh..62..219V,2013SoPh..288..205T, 2015A&A...577A.126M}. These shocks pull the chromospheric materials into the overlying region (Figs. 3-4). Fundamentally, the ponderomotive force is proportional to the gradient of the time-averaged square of the electric field in the multi-fluid plasma species \citep[e.g.,][]{1998PhRvL..80.3523B}. In the present case of single-fluid MHD regime, the ponderomotive force is directly proportional to the gradient of the square of the magnetic field perturbation \citep[e.g.,][]{2016ApJ...829...80B, 1999JPlPh..62..219V}. The ponderomotive force is the negative gradient of the magnetic pressure ($F_{p} = - \nabla (\frac{B^2}{2\mu})$) associated with the transverse Alfv\'enic perturbations.  \\

These spicule-like jets attain their maximum height, thereafter the cool jet plasma falls down under the action of the gravity. These chromospheric materials are moving up and falling down quasi-periodically multiple times within simulation time domain (i.e., 0 to 1000 sec; Fig. 4, lower panel). The jet's velocity arrows (red) are over-plotted on each panel of Fig. 3, which show the velocity of upward propagating and downward falling plasma in the localized solar atmosphere. These Alfv\'en pulse-driven thin spicule-like jets depend on the amplitude of the initial pulses (A$_{v}$). We apply fifteen set of randomly generated amplitudes between 50 and 90 km $s^{-1}$ in the chromosphere at the initial time-span (i.e., t=0 s) of the present simulation. The applied magnitude of the pulses are strong enough to produce spicule-like jets along the field lines. These pulses trigger the spicule-like jets quasi-periodically at different spatio-temporal scales (Fig. 3). Therefore, the amplitude of the initial transverse pulses acts as an initial driver, which plays an important role in the evolution of these spicule-like cool jets. In all fifteen sets of spicule-like jets, we have chosen 3 cases (i.e., case-1, case-2 and case-3) as shown in Fig. 4 (upper-panel) to analyze their evolution/formation, and various physical properties. We track these jets by clicking the path coordinates to estimate the distance-time diagrams in log $\rho$ (Fig. 4, lower-panel), upward velocity ($V_{y}$) (Fig. 5, left-panel), associated amplitude of the $V_{z}$ pulse (Fig. 5, middle-panel), mass flux (Fig. 6) and kinetic energy density (Fig. 7). We estimate the physical parameters along
all three slits (case-1, case-2, case-3) from 2.5 Mm to 8.0 Mm where jets are clearly visible moving along the field lines. The transverse Alfv\'en perturbations (cf., red profile in the right panel of Fig. 5) transfer linearly the energy into the field-aligned magntoacoustic perturbations which later convert into the shocks (cf., blue curves in the right panel of Fig. 5). These shocks lead the chromospheric and transition region plasma following behind them as spicule-like jets in the inner corona quasi-periodically (cf., Fig., 4). They transport the chromospheric mass and energy into the overlying region (Figs 6-7). \\

 The estimated mass flux ($\rho V_{y}$) along spicule-like jets as appeared on the distance time diagrams for all three cases are shown in Fig. 6 (top row). We choose four paths (black-dashed lines) for each case (e.g., case-1, 2, and 3; top row) to estimate the mass fluxes with height. From the first to fourth peak in each distance time diagram, we have demonstrated the mass fluxes associated with the quasi-periodically driven spicule-like jets. The spatial and temporal variation of mass flux associated with first to fourth jet on each representative case (Fig. 6, top-row) clearly show that the quasi-periodic rise and fall of the chromospheric plasma occurs above TR. The estimated mass flux for all three cases (i.e., paths related to case-1, case-2 and case-3, where jets are driven quasi-periodically upward) are of the order of $\simeq$ $ 10^{-6}$ g $cm^{-2}s^{-1}$ (Fig. 6, second to fifth row). However, we notice that mass flux of the first sets of jets (Fig. 6, second row, blue-dotted points fitted with the red curves) show some difference from the second, third and fourth set of jets during their evolution in all three cases (Fig. 6; third-fifth rows). These differences are obviously due to the quasi periodic behaviour of the jets. The first jets in each case (i.e., first chosen paths related to cases 1-3 as shown in top-row of Fig. 6), are most longer and denser jets. They transport large mass flux compared to the other jets rising and falling later on the same path. This significant value of mass flux ($\simeq$ $10^{-6}$ g $cm^{-2} s^{-1}$) exhibits the mass transport in the solar atmosphere during the evolution of spicule-like jets. However, the mass flux as related to the first set of evolved jets (Fig. 6, second row for the paths P11, P21, P31) show that they are produced by the stronger shocks (Fig. 5, right-column). These jets launch a large amount of the chromospheric plasma higher into the inner corona as compared to second set of jets (Fig. 6, third row for the paths P12, P22, P32), third set of jets (Fig. 6, fourth row for the paths P13, P23, P33) and fourth set of jets (Fig. 6, fifth row for the paths P14, P24, P34). It is also demonstrated that the mass flux of the cool plasma quasi-periodically rise and fall along the chosen path (cf., Figs. 8-9; top-row of Fig. 6). \\
 
 The spatial variation of kinetic energy density (KED, in CGS unit) as appeared on the distance time diagram for all three cases (i.e., case-1, case-2 and case-3, where jets rise and fall quasi-periodically) is shown in Fig. 7. We choose the paths (black-dashed lines) which are over plotted on distance-time diagrams of KED for the most longer and denser first sets of spicule-like jets. The KED w.r.t. the height of the spicule-like jets are basically the estimates when they moved upward in the model atmosphere. We do not analyze the KED of these jets during their downfall. We only display the values of KED for the first set of upward moving jets on each chosen path (i.e., second row for P11, P21, and P31 related to cases 1, 2 and 3 respectively; Fig. 7). It can be seen from the distance-time diagrams in the top-panel of Fig. 7 that KEDs are larger compared to the same for the second, third, and fourth set of jets during their evolution. These multiple jets follow the quasi periodic behaviour and KED associated with them is transported from the chromosphere into the inner corona (top-panel in Fig. 7). We find that there is reasonable amount of KED associated with these jets which is transported into upper atmosphere during the evolution of these spicule-like jets. Recently, \citet[][]{2021MNRAS.505...50G} have estimated KEDs for the macrospicules which possess almost similar amount as we estimate in these scale cool spicule-like jets. This trend of KED demonstrate that the energetic chromospheric jets are launched subsequently by the initial action of the Alfv\'en pulses. At the maximum height of the moving jets, the KED and mass flux both are transported upward due to the shock propagation. Firstly the shocks are launched and thereafter the cool plasma always follow in the form of jets (cf., Fig. 8). The first shock is always stronger in all three cases (cf., Fig., 8), which subsequently trigger the most strong jet followed by other jets and related mass motions (Fig. 8). These shocks are strong enough to transport the mass flux and KED in the corona. The spatial distribution of the mass flux and KED along the jet's body during their maximum elongation support the fact that they carry mass and momentum (or energy) into the localized corona (Fig. 6-7). We note that as we increase the magnitude of $V_{z}$ the corresponding $V_{y}$ increases. Therefore the energy related to transverse waves must be transported to the field-aligned motions in the non-linear regime. This also infers that as the amplitude of the velocity of Alfv\'en perturbations increases, the kinetic energy and field-aligned mass flows also increases. This qualitatively infers that the energy is being transferred from Alfv\'en pulses to the spicule-like plasma jets.\\
 
A complete picture emerges when we correlate the upward velocity ($V_{y}$) and mass flux (Fig. 8). Fig. 8 shows the comparison between upward velocity ($V_{y}$) and mass flux for all the three cases. We take horizontal strip at y=3.0 Mm to estimate these parameters (see red horizontal lines in Fig. 6). The variation of mass flux shows its temporal evolution associated with the spicule-like jets crossing a height of 3 Mm and thereafter falling back through the same height along all three considered slits (i.e., case 1, 2 and 3 as shown in Figs. 4, 5). We also choose the horizontal slits at y = 3 Mm on Fig. 5 (left-panels) to estimate the vertical velocity ($V_{y}$) w.r.t. time. The comparison of vertical velocity and mass flux shows that whenever the shock steepens, the spicule is generated and mass flux maximizes at the detection point in the inner corona at y = 3.0 Mm just above the TR (y = 2.7 Mm).\\

Fig. 9 shows the wavelet analysis of the estimated mass flux in the overlying corona at y=3 Mm during the evolution of these jets by using standard tool of wavelet analysis \citep[e.g.,][]{1998BAMS...79...61T}. The temporal variation of the mass flux for case-1, case-2 and case-3 are shown in Fig. 9, upper panel. The lower-left panels of each wavelet diagram show the power wavelet spectrum of the time series. We note that the power lies within the range 3-5 min for case-1, while 2-5 min for case-2, and case-3. We estimate the global power in which the detected periodicities for all the cases are shown in the lower-right panel of each wavelet in Fig. 9 \citep[e.g.,][]{1998BAMS...79...61T}. We calculate the significance level of the global periodicities using power law (green curve in Fig. 9, lower-right panel; e.g., \citealt[][]{2016ApJ...825..110A}).
The black curves in global wavelet show the fast-Fourier transform (FFT) of the time series. The blue curves show the global significance level estimated using white noise model, while the red curves indicate the global significance level estimated using red noise model. The cyan curves in global wavelets are the global wavelet power peaking significantly at 3.9 min. The mass fluxes exhibit $\simeq$ 4 min periodicities with significance level $>$ 95 \% (Fig. 9). The variation of mass flux detected at y=3.0 Mm shows quasi-periodic rise and fall of the plasma in the form of spicule-like jets at time-scale of $\simeq$ 4.0 min. This time-scale is close to the typical scale of 5.0 min periodicity. \\

The rise and fall of the spicules weaken in time as the shock train weakens. This is evident when we construct a time series of the mass flux (Fig. 8 and top panel of wavelets in Fig. 9). We find that the rise of the first spicule carries larger mass flux. Other spicules carry comparatively smaller mass fluxes thereafter. Therefore, the substantial powers in the wavelets are localized in early epoch of the time near the COI. However, a reasonable amount of power also spreads outside the COI in the wavelets. This makes the global power and associated period significant in each wavelet. These global powers peak around a period of $\approx$4.0 min qualifying the significance tests of white, red-noise, and power-law models. These globally significant periodicities depict the time-scale of quasi-periodic rise and fall of the spicule-like jets.

\section{Conclusions}

 We have performed 2.5D ideal MHD simulation to model the evolution of spicule-like jets, which carry the mass and energy into the lower corona. The MHD simulation has been performed by the initial implementation of the velocity pulses in the z-direction (i.e., $V_{z}$) which mimic the transverse Alfv\'en pulses. The action of these transverse perturbations further launches spicule-like cool jets through the open magnetic field lines in the solar atmosphere. Although many authors have reported the triggering mechanisms of these spicule-like jets both in the frame-work of theory and observations \citep[e.g.,][and references cited therein]{1998ESASP.421...19S, 1999ApJ...514..493K,2000SoPh..196...79S, 2004Natur.430..536D, 2017NatSR...743147S, 2019NatCo..10.3504L}, yet it is still debated as one of the open questions in solar physics. These jets have capabilities to transfer the mass and energy into the inner corona (Figs. 6-7). In the present work, we provide a comprehensive numerical simulation and studied kinematics and energetics of the evolved spicule-like cool jets. We find that if the transverse velocity pulse acts on the magnetic field lines in the solar chromosphere with enough strong amplitude (i.e., 50-90 km $s^{-1}$), it produces the field aligned magnetoacoustic perturbations. We implement the transverse pulses randomly at locations in range of 1.5-2.0 Mm height which trigger the chromospheric quasi-periodic jet-like features. These Alfv\'en pulses further transfer energy into the field aligned magnetoacoustic perturbations resulting into the shocks in the stratified atmosphere. This energy exchange can be done in non-linear regime when transverse perturbation is converted into the field aligned motions \citep[e.g.,][]{1998PhRvL..80.3523B, 1999JPlPh..62..219V, 2015A&A...577A.126M, 2016ApJ...829...80B} due to ponderomotive force. These shock trains create lower pressure regions, and cause rise and fall of cool spicule-like jets quasi-periodically. \\

\citet[][]{1982SoPh...75...35H} have reported the formation of solar spicules by implementing Alfv\'en perturbations of 1 km $s^{-1}$ at the photosphere. These velocity perturbations further couple with the magnetoacoustic shocks and launch the transition region plasma into the lower corona in the form of spicules. In our present model, the implemented transverse perturbations generate the Alfv\'en pulses that do not couple with the magnetoacoustic waves. This is due to the fact that in present 2.5D simulation we choose $V_{z}\not=$0 and $B_{z}$=0. These perturbations transfer the energy to the field-aligned magnetoacoustic waves that evolve further into the shocks. This generates the mass motions in form of the spicule-like cool jets in vertical direction. \citet[][]{1999ApJ...514..493K} have also studied the evolution of spicules due to Alfv\'en waves. \citet[][]{2020ApJ...905..168O} provide the analytical model for the generation of spicule-like jets due to the tension component of the Lorentz force. \citet[][]{2008ApJ...683L..83N} have given the physical scenario of the formation of longer chromospheric jets due to Lorentz force and Alfv\'en perturbations at their base. There are also several observational evidences for the generation of spicules and jets due to the evolution of large amplitude Alfv\'en waves in the chromosphere \citep[e.g.,][and references cited therein]{2007Sci...318.1574D, 2011Natur.475..477M, 2017NatSR...743147S}. \\

It is proposed by several authors that the Alfv\'en perturbations constituting the incompressible transverse waves can be produced by the granular motions acting on the footpoint of the flux-tubes/coronal loops anchored at the solar photosphere, or at the large spatial scales by the turbulent convective motions churning out there \citep[e.g.,][and references cited there]{1981SoPh...70...25H, 1982ApJ...257..345H, 1999ApJ...514..493K, 2009Sci...323.1582J, 2010ApJ...710.1857M, 2021JGRA..12629097S}. Such perturbations can also be generated by the swirling/rotatory motions of the magneto-plasma systems locally, and by torsional Alfv\'en waves propagating up in the overlying atmosphere \citep[e.g.,][and references cited there]{2017NatSR...743147S, 2019NatCo..10.3504L}. Sometime, the mode conversion of the magnetoacoustic perturbations generated at the photosphere further lead the formation of the Alfv\'en waves in the chromosphere where plasma beta becomes unity \citep[e.g.,][and references cited there]{2010CSci...98..295D, 2018NatPh..14..480G, 2021JGRA..12629097S}. In the present work, we invoke the physical scenario that dynamical chromosphere may launch randomly the transverse velocity perturbations akin to the Alfv\'en pulses and subsequent interaction with the plasma may take place. These pulses transfers energy to the field aligned perturbations to launch the magnetoacoustic shocks and spicule-like jets. We generate the obtained physical scenario by considering $B_{z}$ = 0 and putting $V_{z}$ pulse on the strong magnetic fields $B=B_{e}(x,y)$, that generates pure Alfv\'en pulses. These perturbations cause the evolution of magnetoacoustic waves by the ponderomotive force, which further steepens into the shocks. The modeled multiple jets are evolved along the magnetic field lines and transport the mass above the transition region into the inner corona. \\ 

The present simulation is performed in the ideal limits of magnetohydrodynamics. However, the non-ideal plasma conditions may affect the evolution of various MHD waves ans associated plasma dynamics. Recently, the slow magnetoacoustic waves were found to be affected by the thermal misbalance between heating and cooling processes of the plasma \citep[e.g.,][and references cited there]{2017ApJ...849...62N, 2019A&A...628A.133K, 2020A&A...644A..33K, 2020SoPh..295..160B, 2021SoPh..296...98B, 2021SoPh..296..122B, 2021SoPh..296...20P, 2021SoPh..296..110P}. \citet{2021SoPh..296..122B} have recently reported that Alfv\'en waves can trigger the longitudinal motions of the confined plasma under the effect of heating-cooling misbalance. They showed analytically that the heating cooling misbalance causes an exponential bulk plasma flows, which themselves alter the Alfv\'en-induced plasma motions. The periodic Alfv\'en wave frequency crucially determines the amplitude and phase shift of the induced longitudinal motions. In the present simulation, we do not consider the inclusion of the heating-cooling imbalance on the formation of the spicule-like jets. However, it will be an outstanding science case to be taken at the next step in future to understand numerically the effect of such physical process on the formation of the quasi-periodic plasma motions. \\

It is also well observed in the past that the spicules are launched due to the leakage of the p-mode powers subsequent and formation of the associated shocks \citep[e.g.,][]{2004Natur.430..536D}. However, the present simulation demonstrates that under the given physical scenario, the quasi-periodic mass motions at typical $\simeq$ 4.0 min time-scale can be generated $\it in$ $situ$ coupling the chromosphere-TR-inner corona. These quasi-periodic mass motions (typical rise and fall of the cool jets) are seen in the observations also \citep[e.g.,][]{2014Ap&SS.354..259K, 2015ApJ...815L..16S}. Such periodic/quasi-periodic rise and fall of the cool jets are physically well demonstrated in the present numerical simulations. In particular, the detected period of rise and fall of mass in the present case match well with the observed period reported by \citet[]{2014Ap&SS.354..259K}. There are also several other physical scenarios reported, which claim the formation of the spicule-like jets due to direct onset of the Lorentz force \citep[e.g.,][and references cited therein]{2008ApJ...683L..83N, 2017Sci...356.1269M, 2017ApJ...848...38I, 2020ApJ...905..168O}. However, the present model provides the physical inference of the indirect role of such transverse perturbations implemented in terms of velocity pulses in the solar chromosphere. It demonstrates the significance of subsequent generation of the magnetoacoustic shocks and thereby quasi-periodic rise and fall of the spicule-like cool jets. The 4-m DKIST observations may shed new light on such conversions at finer spatial scales in the localized lower solar atmosphere \citep{2021SoPh..296...70R}. As demonstrated, these plasma dynamics are efficient in transporting mass and energy upward up to the inner corona. They partially contribute to the energy and mass transport, therefore, in the heating of the localized corona.\\

\section*{Acknowledgements}

We thank reviewer for his/her significant scientific suggestions that improved manuscript considerably. AKS thanks and acknowledge Prof. K. Murawski for valuable discussions and realistic solar atmosphere model. B.S. thankfully acknowledge the Human Resource Development Group (HRDG), Council of Scientific $\&$ Industrial research (CSIR) India for providing him a senior research scholar grant. Authors acknowledge the use of PLUTO code and of PYTHON libraries for numerical data analysis.  B.S. also acknowledge the help of Ms. Kartika Sangal in the wavelet analysis.

\section*{Data Availability}
The data underlying this article will be shared on reasonable request to the corresponding author.











\bsp	
\label{lastpage}
\end{document}